\documentstyle[12pt,epsfig]{article}             
\newcommand{\beq}{\begin{equation}}             
\newcommand{\eeq}{\end{equation}}               
\newcommand{\bqry}{\begin{eqnarray}}            
\newcommand{\eqry}{\end{eqnarray}}              
\newcommand{\bqryn}{\begin{eqnarray*}}          
\newcommand{\eqryn}{\end{eqnarray*}}            
\newcommand{\preprint}[1]{\begin{table}[t]      
            \begin{flushright}                  
            \begin{large}{#1}\end{large}        
            \end{flushright}                    
            \end{table}}                        
\newcommand{\PD}[2]                             
    {\frac{\partial^{#2}}{\partial #1^{#2}}}    
\begin{document}
\preprint{LA-UR-99-5914}
\title{Melting as a String-Mediated Phase Transition}
\author{\\ Leonid Burakovsky\thanks{E-mail: BURAKOV@T5.LANL.GOV}, \
Dean L. Preston\thanks{E-mail: DEAN@LANL.GOV}, \
and Richard R. Silbar\thanks{E-mail: SILBAR@WHISTLESOFT.COM. Also at 
WhistleSoft, Inc., 168 Dos Brazos, Los Alamos, \hspace*{0.5cm} NM 87544, USA}
 \\  \\ 
Los Alamos National Laboratory \\ Los Alamos, NM 87545, USA}
\date{ }
\maketitle
\begin{abstract}
We present a theory of the melting of elemental solids as a 
dislocation-mediated phase transition. We model dislocations near melt as 
non-interacting closed strings on a lattice. In this framework we derive 
simple expressions for the melting temperature and latent heat of fusion 
that depend on the dislocation density at melt. We use experimental data 
for more than half the elements in the Periodic Table to determine the 
dislocation density from both relations. Melting temperatures yield a 
dislocation density of $(0.61\pm 0.20)\;\!b^{-2},$ in good agreement with 
the density obtained from latent heats, $(0.66\pm 0.11)\;\!b^{-2},$ where 
$b$ is the length of the smallest perfect-dislocation Burgers vector. 
Melting corresponds to the situation where, on average, half of the atoms 
are within a dislocation core.
\end{abstract}
\bigskip
{\it Key words:} melting, dislocation, phase transition, lattice 

PACS: 11.10.Lm, 11.27.+d, 61.72.Bb, 64.70.Dv, 64.90.+b 
\bigskip

\section{Introduction}

Nearly 50 years ago Shockley \cite{Shock} successfully accounted for the 
fluidity of a liquid by assuming a certain concentration of line defects in 
the liquid state. Bragg \cite{Bragg} had earlier estimated an upper bound on 
the core energy of a dislocation under the assumption that the core atomic 
configuration was like that of a liquid. Cotterill and Doyama \cite{CD} later 
confirmed the Bragg estimate. These early results implied that the liquid 
state is equivalent to a crystal saturated with dislocation cores. It was 
first suggested by Mott \cite{Mott} that the melting transition could be 
described in terms of dislocations. 

There are now compelling results from molecular dynamics \cite{MD} and 
Monte Carlo \cite{MC} simulations that imply that dislocations play a key role 
in three-dimensional melting, and moreover there is experimental evidence that 
linear defects are in fact generated near the melting transition \cite{Crawf}. 

Mizushima \cite{Miz} and Ookawa \cite{Ook} were the first to formulate a 
dislocation theory of melting. They based their theory on the fact that the 
self-energy of a dislocation decreases with dislocation density because of 
screening. In their theory melting is a first-order transition that occurs 
when the free energy of the crystal with a sufficiently high concentration 
of thermally-generated dislocations equals the free energy of the 
dislocation-free crystal. Their predicted melting temperatures agree with 
data for reasonable choices of the core energy. Many dislocation theories 
of melting then followed [10$\;\!$-13], most notably the exhaustive 
treatment of linear-defect-mediated melting by Kleinert and collaborators 
\cite{KleinII}. We refer the reader to several fine reviews of the literature 
for additional details and references on dislocation-mediated melting 
\cite{Cotterill,reviews}. 

Significant progress in our understanding of melting has been achieved by 
Kleinert \cite{KleinII} who pointed out that the melting process cannot proceed
through the mediation of dislocations alone. Dislocations are associated with 
the discrete translational symmetry of the crystal, so only this symmetry is 
lost when dislocations condense. But the rotational order of the solid is also 
lost as the solid converts into liquid, and for this to occur the defects 
associated with the rotational symmetry of the lattice, namely disclinations, 
must come into play. Kleinert assumes that the free energy of dislocations 
alone would lead to a second-order phase transition, and in addition shows that
a second-order proliferation of disclinations can occur in a background of 
dislocations above some critical density. The coupled dislocation-disclination 
system could, however, undergo a first-order transition, i.e., melting. 

Although disclinations must participate in the melting process, in this paper 
we consider only the dislocation degrees of freedom. Ideally, we would derive 
the precise form of the free energy of a dense ensemble of dislocations 
interacting via the full Blin potential \cite{KleinII} and subject to specific 
configurational constraints (Brownian, self-avoiding, etc.), but this problem 
has so far defied solution. Instead we develop an effective theory of melting 
based on perfectly screened, non-interacting dislocations. We employ the 
widely accepted $-\rho \ln \rho $ form $(\rho $ is dislocation density) for 
the self-energy density of dislocations \cite{Miz,YI,KV}, which results in 
a first-order phase transition. A dislocation in a dense ensemble of other 
dislocations is assumed to be a random loop, i.e., the possible configurations 
of a dislocation loop are closed random walks, and short-range steric 
interactions are neglected. Thus the partition function is evaluated in 
the independent-loop approximation. We obtain two new relations: a simple 
expression for the melting temperature (the melting relation) that explicitly 
takes into account the crystal structure, and another relation between melting 
temperature, latent heat of fusion, and critical density of dislocations. We 
carry out a comprehensive comparison of these relations with experimental data 
on over half of the Periodic Table. The melting relation is accurate to 17\% 
\cite{previous}. Dislocation densities as determined from the melting 
temperature and latent heat relations are $\rho =(0.61\pm 0.20)\;\!b^{-2}$ 
and $(0.66\pm 0.11)\;\!b^{-2},$ respectively, where $b$ is the length of the 
smallest perfect-dislocation Burgers vector. Both relations should also 
apply to alloys and compounds. 

In Section 2 we discuss the statistical mechanics of dislocation loops on a 
lattice. These results are used to derive the melting relation in Section 3, 
the free energy density in Section 4, and the formula for the latent heat of 
fusion in Section 5. The values for the critical dislocation density extracted 
from both the melting relation and the formula for the latent heat of fusion   
are checked in Section 6 with the formula for volume change at melt. Our 
concluding remarks appear in Section 7. 

\section{Statistical mechanics of dislocation loops on a lattice}

The energy per unit length, $\sigma ,$ of a dislocation 
can be very large. However, this energy can always be 
compensated at sufficiently high temperatures by the large entropy of 
line-like structures, as will be seen in what follows. 

In a Bravais lattice with coordination number $z$ we consider the graph 
$\Gamma ,$ the edges of which are all nearest-neighbor links. The set 
of $z$ links from any lattice site is identical to the set of shortest 
perfect-dislocation Burgers vectors, of length $b.$ We now evaluate the 
partition function for a single Brownian loop on $\Gamma .$

The line tension, $\sigma ,$ is assumed to be independent of its length, 
$L,$ as is the case for a dislocation in a dense complex network. (In a 
dilute network, interactions between distant segments of a dislocation lead 
to a logarithmic dependence of $\sigma $ on $L.)$ The number of configurations 
of a string of length $L$ is $(z')^{L/b},$ where $z'$ is the number of 
possible directions that a line segment can take from a given lattice site. 
If backtracking is not allowed, $z'=z-1.$ For a simple cubic lattice in $D$ 
dimensions $z=2D.$
 
Hence, the partition function for a single closed dislocation 
(in 3 dimensions) is 
\beq
Z_1=\sum _Lp(L,V)(z')^{L/b}e^{-\beta \sigma L}=
\sum _Lp(L,V)e^{-\beta \sigma _{{\rm eff}}L},\;\;\;\sigma _{{\rm eff}}\equiv 
\sigma \left( 1-\frac{T\ln z'}{\sigma b}\right) ,
\eeq
where $\beta \equiv 1/k_BT,$ $p(L,V)$ is the sum of probabilities over all 
lattice sites that a dislocation of length $L$ will close, $V$ is the volume 
of the system, and $\sigma _{{\rm eff}}$ is the {\it effective} energy cost 
to create unit length of a string at temperature $T.$ 

In order to calculate $p(L,V),$ let $p({\bf r}',{\bf r};L/b)$ be the 
probability density for a dislocation of $L/b=n$ steps to start at ${\bf r}$ 
and end at ${\bf r}'.$ In the limit $n\rightarrow \infty ,$ $b\rightarrow 0,$ 
and $L={\rm const,}$ $p({\bf r}',{\bf r};L/b)$ satisfies the diffusion 
equation \cite{Wie} 
\beq
\frac{\partial p}{\partial n}=\frac{b^2}{z'}\nabla ^2p,
\eeq
the solution of which is the heat-kernel expansion
\beq
p({\bf r}',{\bf r};L/b)\equiv p({\bf r}'-{\bf r};L/b)=\sum _{{\bf k}}f_{\bf k}
({\bf r})f^\ast _{\bf k}({\bf r}')e^{-E_{{\bf k}}L/b},
\eeq
where the $f_{{\bf k}}({\bf r})$ are eigenfunctions of the Laplacian,
\beq
-\frac{b^2}{z'}\nabla ^2f_{{\bf k}}=E_{{\bf k}}f_{{\bf k}},
\eeq
which we take as normalized according to
\beq
\int d^3{\bf r}\;\!|f_{{\bf k}}({\bf r})|^2=1.
\eeq
When $V\rightarrow \infty ,$ we have
\beq
f_{{\bf k}}({\bf r})=\frac{1}{\sqrt{V}}\;\!e^{i{\bf k}\cdot {\bf r}},\;\;\;
E_{{\bf k}}=\frac{b^2k^2}{z'},
\eeq
where $0\leq k=|{\bf k}|\leq \infty .$ It then follows from Eq.\ (3), in 
which we replace a sum by an integral, $\sum _{\bf k}\rightarrow V/(2\pi )^3
\int d^3{\bf k},$ that
\beq
p({\bf r}'-{\bf r};L/b)=\int \frac{d^3{\bf k}}{(2\pi )^3}\;\!e^{i{\bf k}\cdot 
({\bf r}-{\bf r}')-bLk^2/z'}=\left( \frac{z'}{4\pi bL}\right) ^{3/2}
e^{-z'({\bf r}'-{\bf r})^2/4bL}.
\eeq
The normalization (5) of the eigenfunctions thus imparts unit  
normalization to the probability density:
\beq
\int d^3{\bf u}\;p({\bf u};L/b)=1.
\eeq
The partition function for a dislocation loop $({\bf r}'={\bf r})$ is 
therefore 
\beq
Z_1=\sum _L\frac{b}{L}\int d^3{\bf r}\;p({\bf 0};L/b)e^{-\beta \sigma _{{
\rm eff}}L}=\left( \frac{z'}{4\pi }\right) ^{3/2}\frac{V}{b^3}\sum _L\left( 
\frac{L}{b}\right) ^{-5/2}\!e^{-\beta \sigma _{{\rm eff}}L}\equiv \sum _LN(L)
e^{-\beta \sigma L},
\eeq
where the factor $b/L$ removes the overcounting due to the degeneracy in 
the number of starting points on the loop. Here, $N(L)$ is the number of 
configurations of a loop of length $L.$ The exponent 5/2 becomes $1+D/2$ in 
$D$ dimensions \cite{GSW}. 

Real dislocations are not necessarily Brownian loops. In fact, they are 
expected to be self-avoiding and/or neighbor-avoiding loops, so they do not 
penetrate each other's core. Eq.\ (9) can then be extended to non-Brownian 
or open dislocations by means of an effective exponent $q+1\neq 5/2$ and 
normalization constant $A(q,z')$ \cite{Copeland}, as follows:
\beq
Z_1=A(q,z')\;\!\frac{V}{b^3}\sum _L\left( \frac{L}{b}\right) ^{-q-1}\!\!
e^{-\beta \sigma _{{\rm eff}}L}.
\eeq
Here, $q=-1$ for non-interacting (Brownian) open dislocations \cite{Copeland} 
and $q\approx 7/4$ for self-avoiding dislocations at low densities in 3 
dimensions \cite{Copeland}. In the string literature, the value $q=0$ has also 
been quoted. A general argument based on modular invariance \cite{BV} shows 
that for non-interacting closed strings $q=0$ for sufficiently high energy 
on any compact target space. The same value of $q$ was also obtained in a 
discrete model for strings \cite{SS}, and as a static solution to the 
string Boltzmann equation \cite{LT}. In principle, $q$ may even be a 
function of temperature. Although we may expect $3/2\leq q\leq 7/4$ 
\cite{Copeland2}, our main conclusions do not depend on the precise value 
of $q.$ The normalization constant, $A(q,z'),$ can be calculated analytically 
for Brownian loops in any dimension $(q=D/2),$ analogously to the calculation 
of $A(3/2,z')=(z'/4\pi )^{3/2}$ in Eq.\ (9), and numerically in other cases. 

The average length of a loop is 
\beq
\langle L\rangle =\frac{\sum _LLN(L)e^{-\beta \sigma L}}{\sum _LN(L)e^{-\beta 
\sigma L}}=\frac{\xi (T)}{\bar{\xi }(T)}\;\!b, 
\eeq
where we define 
\beq
\xi (T)\equiv A(q,z')\sum _{L/b}\left( \frac{L}{b}\right) ^{-q}e^{-\beta 
\sigma _{{\rm eff}}L}, 
\eeq
and 
\beq
\bar{\xi }(T)\equiv A(q,z')\sum _{L/b}\left( \frac{L}{b}\right) ^{-q-1}
e^{-\beta \sigma _{{\rm eff}}L}=\frac{b^3}{V}\;\!Z_1. 
\eeq

The grand canonical partition function for an ensemble of non-interacting 
indistinguishable loops is given by 
\beq
Z=Z(T,V,\mu )=\sum _{N=1}^\infty \frac{Z_1^N}{N!}\;\!e^{\mu N/k_BT}=
\exp \;\!\{ \exp \left( \frac{\mu }{k_BT}\right) Z_1\}, 
\eeq
where $\mu $ is the chemical potential. The free energy of the ensemble is
\beq
F=-k_BT\ln Z=-k_BTe^{\mu /k_BT}Z_1.
\eeq
The average number of loops in the ensemble is
\beq
\bar{N}=-\left( \frac{\partial F}{\partial \mu }\right) _{T,V}\!\!\!=
e^{\mu /K_BT}Z_1. 
\eeq

Since $Z=\sum _{L_i}N(L_i)e^{-\beta (\sigma L_i-\mu )},$ where $N(L_i)$ is 
the number of dislocation configurations of total length $L_i,$ the average 
total dislocation length in the ensemble is
$$\bar{L}=\frac{1}{Z}\;\!\sum _{L_i}L_iN(L_i)e^{-\beta (\sigma L_i-\mu )}=
-\frac{\partial \ln Z}{\partial (\beta \sigma )}=-e^{\mu /k_BT}\frac{\partial 
Z_1}{\partial (\beta \sigma )}$$
\beq
=\bar{N}\;\!\frac{\sum _LLN(L)e^{-\beta \sigma L}}{\sum _LN(L)e^{-\beta \sigma
L}}=\bar{N}\langle L\rangle ,
\eeq
i.e., the average total dislocation length is equal to the average number of 
loops times the average loop length. 

The dislocation density, $\rho ,$ is the average total length per 
unit volume. It then follows from (12),(13) and (17) that  
\beq
b^2\rho (T)=\frac{\bar{N}\langle L\rangle }{V}\;\!b^2=\frac{\bar{N}}{V}
\;\!\frac{\xi (T)}{\bar{\xi }(T)}\;\!b^3=e^{\mu /k_BT}\xi (T). 
\eeq

\section{New melting relation}

The effective line tension, $\sigma _{{\rm eff}},$ [see Eq.\ (1)] vanishes 
at the critical temperature $k_BT_{cr}\equiv \sigma b/\ln z'.$ Consequently, 
dislocations proliferate as $T_{cr}$ is approached from below. At temperatures 
above $T_{cr},$ the divergence of $Z_1$ signals the breakdown of the 
underlying theory, and the system enters a new phase. Hence, the temperature 
$T_{cr}$ corresponds to a phase transition, in which dislocations are 
copiously produced in the solid. We therefore equate the melting temperature, 
$T_m,$ to $T_{cr}.$ 

The line tension, i.e., the dislocation self-energy per unit length, is 
assumed to be that of a dislocation in a complex array, or tangle, of other 
dislocations. In that case the stress field of a given dislocation beyond 
$\simeq R/2,$ where $R$ is the mean interdislocation separation, is largely 
cancelled out by the stress fields of the other dislocations in the complex 
array \cite{HL,Sar}. The line tension is then the sum of the core energy 
plus the elastic energy inside a cylinder of radius $R/2$ \cite{HL}: 
\beq 
\sigma =\kappa \frac{Gb^2}{4\pi }\ln \left( \frac{\alpha }{b}\frac{R}{2}
\right) =\kappa \frac{Gb^2}{8\pi }\ln \left( \frac{\alpha ^2}{4b^2\rho }
\right) . 
\eeq 
Here, $\kappa $ is 1 for a screw dislocation and $(1-\nu )^{-1}\approx 3/2$ 
for an edge dislocation, $\nu $ being the Poisson ratio. Also, $G$ is the 
shear modulus, $b$ is the Burgers vector magnitude, and $\alpha $ is 
a constant of order unity. In the second half of this equation we have taken 
distance $R$ to be approximately equal to $1/\sqrt{\rho },$ where $\rho $ is 
the dislocation density defined in Eq.\ (18). An expression of the form (19) 
with $R=\rho ^{-1/2}$ for the dislocation self-energy was originally proposed 
by Mizushima \cite{Miz}, later put on a sound theoretical basis by Yamamoto 
and Izuyama \cite{YI}, and was recently employed by Kierfeld and Vinokur 
\cite{KV} to model dislocation-mediated phase transitions of vortex-line 
lattices in high-$T_c$ superconductors.  

The constant $\alpha $ accounts for the nonlinear elastic effects in the 
dislocation core. Hirth and Lothe \cite{HL} compare dislocation energies in 
the Peierls-Nabarro (discrete) and Volterra (continuum) dislocation models and
find
\beq
\frac{1}{\alpha }=\frac{d}{eb}\left( \frac{\sin ^2\beta }{e^\gamma (1-\nu )}+
\cos ^2\beta \right) ,
\eeq
where $\gamma =(1-2\nu )/4(1-\nu )\approx 1/8,$ $d$ is the interplanar 
spacing, and $\beta $ is the angle between the Burgers and sense vectors of 
the dislocation. In a face-centered cubic (fcc) crystal, the smallest 
perfect-dislocation Burgers vectors are $\frac{1}{2}\langle 110\rangle a,$ and 
the primary glide planes are $\{111\}$ with $d=a/\sqrt{3},$ where $a$ is the 
lattice constant. Experimental evidence (ref. \cite{HL}, Table 9-2, p.\ 275) 
suggests that the predominant high-temperature glide system in body-centered 
cubic (bcc) lattices is $\{110\},$ which has $d=a/\sqrt{2}.$ The smallest bcc 
perfect-dislocation Burgers vectors are $\frac{1}{2}\langle 111\rangle a.$ 
Thus, in both cases $d/b=\sqrt{2/3}.$ Averaging over $\beta ,$ 
we find $\alpha \approx 2.9$ for {\it both} fcc and bcc lattices. Atomistic 
calculations of core energies in ionic crystals (ref. \cite{HL}, p.\ 232) 
indicate that $\alpha \approx 3.$ In metals, no such calculations have been 
performed. We use $\alpha =2.9$ for all elements. 

We have also assumed that no backtracking is allowed for dislocations, 
$z'=z-1,$ since each backtracking would result in a divergence in the linear 
elastic interaction energy between the overlapping segments. The coordination 
numbers for the elements considered in our analysis below are $z=6$ for a 
simple cubic (sc) lattice, $z=8$ for bcc and body-centered tetragonal (bct) 
lattices, and $z=12$ for fcc, hexagonal close-packed (hcp), and double hcp 
(dhcp) lattices. Replacing $$b^3\equiv \lambda v_{WS},$$ where $v_{WS}$ is 
the volume of the Wigner-Seitz cell of the crystal lattice and $\lambda $ 
is a geometric constant, we finally obtain our formula for the melting 
temperature of the elements:
\beq
T_m=\frac{\lambda Gv_{WS}}{4\pi \delta \ln (z-1)},\;\;\;
\delta ^{-1}\equiv \kappa \ln \left( \frac{1.45}{b\sqrt{\rho (T_m)}}\right) .  
\eeq   

In ref.\ \cite{previous} we evaluated $Gv_{WS}/4\pi T_m\ln (z-1)$ for 51 
elements and found $\delta /\lambda $ to be $1.01\pm 0.17,$ where the error 
is the root-mean-square deviation. These $\delta /\lambda $ are summarized in 
Fig. 1.

\begin{center}
\vspace{1.5cm} 
\epsfig{file=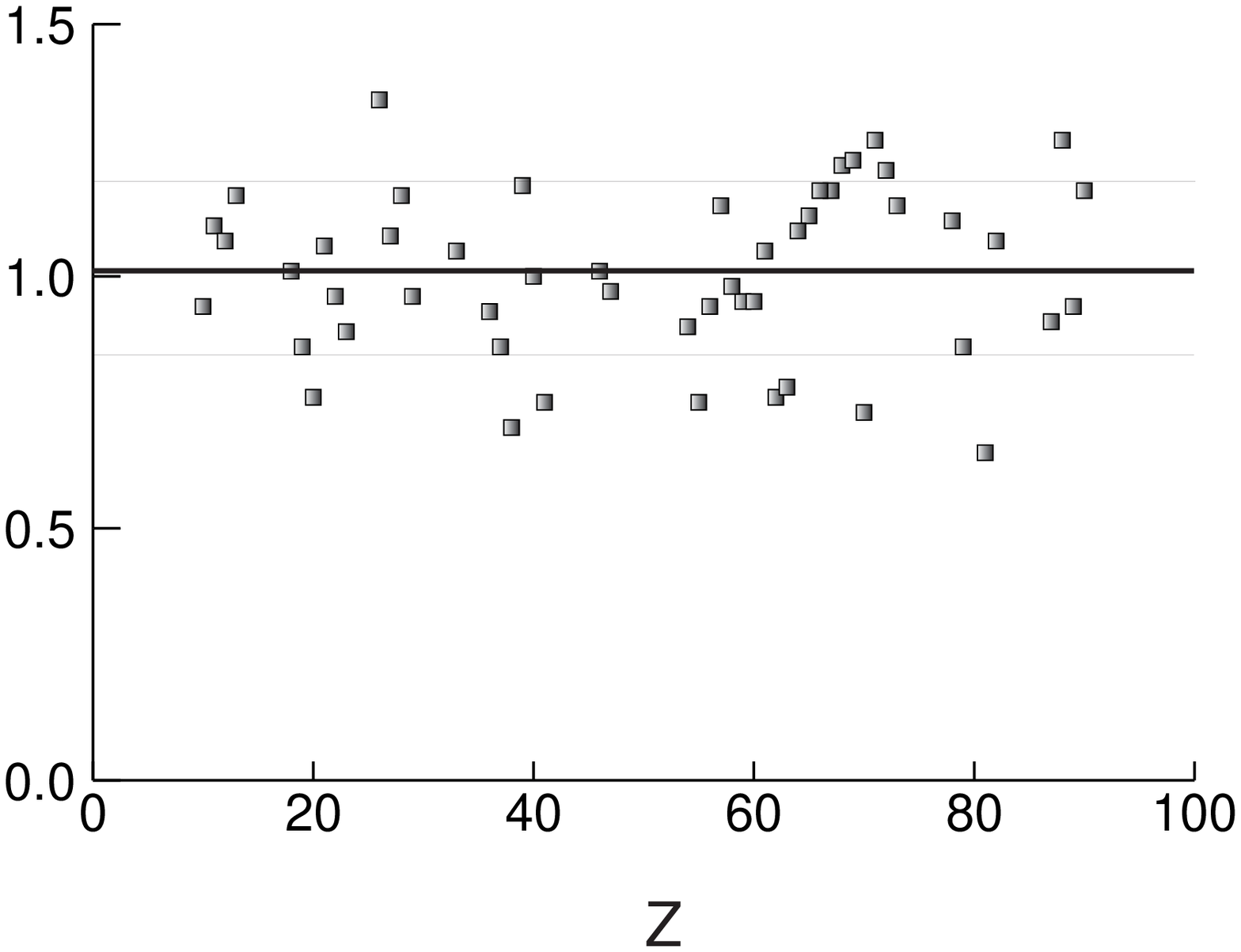,width=15cm,angle=0}
\end{center}
Fig.\ 1. Values of $\delta /\lambda =Gv_{WS}/4\pi T_m\ln (z-1)$ from 
experimental data for 51 elements.
 \\

We now assume that the dislocation ensemble is dominated by perfect 
dislocations with the smallest possible Burgers vectors, since dislocation 
energy is proportional to $b^2.$ 
For a bcc crystal $b=a\sqrt{3}/2,$ $v_{WS}=a^3/2,$ and $b=a/\sqrt{2},$ $v_{WS}
=a^3/4$ for a fcc crystal. Then, $b^3\approx 1.30\;\!v_{WS}$ and $1.41\;\!v_{
WS},$ respectively. For a hcp lattice, $\lambda =(4/\sqrt{3})\;\!(c/a)^{-1},$ 
so that for an ideal hcp crystal $(c/a=\sqrt{8/3})$ one would have $\lambda =
\sqrt{2}.$ As estimated for two hcp metals \cite{PP}, $b^3\approx 1.42\;\!v_{
WS}$ for Mg and $1.24\;\!v_{WS}$ for Zn. Hence, we take $\lambda =1.33\pm 0.09
\approx 4/3.$ This embraces all of the values quoted above. 

In an ensemble of loops there are roughly equal amounts of edge and screw 
dislocation in the crystal, so we have $1/\kappa =(1-\nu /2)\pm \nu/2\simeq 
5/6\pm 1/6.$ Therefore, as follows from (21),
\beq
\ln \left( \frac{2.1}{b^2\rho (T_m)}\right) =\frac{2\;\!(5/6\pm 1/6)}{
(1.33\pm 0.09)(1.01\pm 0.17)}=1.24\pm 0.33.
\eeq
Hence, 
\beq
\rho (T_m)=(0.61\pm 0.20)\;\!b^{-2}.
\eeq 

It follows from Eqs.\ (21)-(23), with $\kappa \lambda =1.6\pm 0.3,$ that 
to $\sim 20$\% accuracy 
\beq
T_m=\frac{Gv_{WS}}{4\pi \ln (z-1)}. 
\eeq
We regard Eq.\ (24) as a new dislocation melting law. 

\section{Free energy of the dislocation ensemble}

To calculate the free energy of a dislocation ensemble, Eq.\ (15), let 
us rewrite Eqs.\ (10) and (12), using Eq.\ (19) with $R=1/\sqrt{\rho },$ 
and replace the sums (which start with $L=4b,$ the smallest loop length) 
by the corresponding integrals:
\beq 
Z_1=\frac{V}{b^3}\;\!A(q,z')\int _4^\infty dx\;\!x^{-q-1}\left[ \frac{4b^2\rho 
}{\alpha ^2}\;\!(z')^{1/c}\right] ^{cx}\!, 
\eeq
\beq 
\xi (T)=A(q,z')\int _4^\infty dx\;\!x^{-q}\left[ \frac{4b^2\rho }{\alpha ^2}
\;\!(z')^{1/c}\right] ^{cx}\!,\;\;\;c\equiv \frac{\kappa Gb^3}{8\pi k_BT}.  
\eeq   
Here, $(4b^2\rho /\alpha ^2)\;\!(z')^{1/c}=\exp \;\!\{-8\pi \sigma 
_{{\rm eff}}/\kappa Gb^2\}\leq 1,$ since $\sigma _{{\rm eff}}\geq 0.$ 
Integrating Eq.\ (25) by parts we find
$$Z_1=\frac{V}{qb^3}\left( \!\frac{\kappa Gb^3}{8\pi k_BT}\ln \left[ \frac{
4b^2\rho }{\alpha ^2}\;\!(z')^{1/c}\right] \!\cdot \xi (T)+\frac{A(q,z')}{4^q}
\left[ \frac{4b^2\rho }{\alpha ^2}\;\!(z')^{1/c}\right] ^{4c}\right) $$
\beq
=\frac{V}{qb^3}\left[ -\frac{\sigma _{{\rm eff}}b}{k_BT}\;\!\xi (T)+\frac{
A(q,z')}{4^q}\;\!e^{-4\sigma _{{\rm eff}}b/k_BT}\right] .
\eeq
Hence,
\beq
F=-k_BTe^{\mu /k_BT}Z_1=
\frac{V}{qb^3}\left( \sigma _{{\rm eff}}\rho \;\!b^3-\frac{A(q,z')e^{
\mu /k_BT}}{4^q}k_BTe^{-4\sigma _{{\rm eff}}b/k_BT}\right) ,
\eeq
where we have replaced $(\kappa Gb^3/8\pi )\;\!\ln(\alpha ^2/4b^2\rho )-k_BT
\ln z'$ by $b\sigma _{{\rm eff}},$ in view of (1),(19), and used Eq.\ (18). 

The second term on the right-hand side of Eq.\ (28) takes its largest value at 
$T=T_m,$ where $\sigma _{{\rm eff}}=0.$ To estimate its contribution to the 
free energy, consider the case of Cu discussed in more detail below. In this 
case, to estimate $A(q,z')\exp\{\mu (T_m)/k_BT_m\}/4^q,$ we use Eqs.\ 
(12),(18), and replace the sum by an integral: 
$$\frac{A(q,z')e^{\mu (T_m)/k_BT_m}}{4^q}=\frac{A(q,z')}{4^q}\;\!\frac{b^2\rho 
(T_m)}{\xi (T_m)}=\frac{b^2\rho (T_m)}{4^q\int _4^\infty dx\;\!x^{-q}}=\frac{
b^2\rho (T_m)\;\!(q-1)}{4}.$$ 
As discussed in Section 2, the value of $q$ may be expected between 
3/2 (Brownian loops) and $\approx 7/4$ (self-avoiding loops). 
With $b^2\rho (T_m)$ given in Eq.\ (23), we therefore obtain  
$A(q,z')\exp\{\mu (T_m)/k_BT_m\}/4^q=0.095\pm 0.037\approx 0.1.$ 

Hence, the contribution of the second term to $qF/V$ would be $\simeq -0.7$ 
meV $\stackrel{\circ }{{\rm A}}$$^{-3}.$ As seen in Fig.\ 2, this contribution 
is negligibly small. In fact, the second zero of $F$ for $T=T_m$ and $A(q,z')
\exp\{\mu (T_m)/k_BT_m\}/4^q=0.1$ occurs at $b^2\rho =0.61,$ which is within 
$\sim 5$\% of the value of 0.64 (the second zero of $F$ at $T=T_m$ with $F/V$ 
given in (29)), and within uncertainties in the values of $b^2\rho (T_m)$ in 
Eq.\ (23).  

Thus, we have derived the dislocation free energy density, and it is given 
approximately by 
\beq
\frac{qF(\rho )}{V}\simeq \sigma _{{\rm eff}}\;\!\rho =-\left( \frac{\kappa 
G}{8\pi }\ln \left( \frac{4b^2\rho  }{\alpha ^2}\right) +\frac{k_BT}{b^3}
\ln (z-1)\right) b^2\rho .
\eeq
This form for the free energy density was previously suggested but not 
derived by Cotterill \cite{Cotterill}. It was later put on a firm theoretical 
basis by Yamamoto and Izuyama \cite{YI}. It also is a fundamental ingredient 
in the recently developed theory of dislocation-mediated phase transitions 
of vortex-line lattices in high-$T_c$ superconductors \cite{KV}.  

In Fig.\ 2 we plot $qF(\rho )/V$ from (29) for Cu for three different 
temperatures: $T<T_m,$ $T=T_m$ and $T>T_m.$ We take $\kappa =6/5,$ $\alpha =
2.9,$ $G=47.7$ GPa \cite{GS}, $T_m=1356$ K, and $b=2.55$ $\stackrel{\circ }{
{\rm A}}$ \cite{previous}. A first-order phase transition, that is melting, 
takes place when the second zero of $F(\rho )$ occurs at the critical 
dislocation density, $\rho (T_m).$ This is a transition from a perfect 
crystalline solid to a highly dislocated {\it solid,} not a liquid. In fact, 
our theory describes dislocations, which do not exist in liquids. If a 
dislocation is viewed as a disclination dipole \cite{dipole}, the dislocated 
solid may in turn undergo a Kosterlitz-Thouless-like transition \cite{KT} to 
a phase of free disclinations, i.e., a liquid. This dislocated solid may then 
be viewed as the three-dimensional analog of an intermediate hexatic phase, 
between a solid and a liquid, in the Halperin-Nelson theory of two-dimensional 
melting \cite{HN}. The clarification of this point needs further 
investigation, to be undertaken elsewhere. Patashinskii {\it et al.} 
\cite{Pat} also identified melting as a transition from a perfect crystalline 
solid to a highly dislocated solid, and Nelson and Toner \cite{NT} found 
residual bond-orientational order in a three-dimensional solid with an 
equilibrium concentration of unbound dislocation loops, which is analogous 
to that in the two-dimensional hexatic phase. 

Note that it is not possible to increase the dislocation density progressively
from zero to $\rho (T_m)$ at a temperature lower than $T_m$ (e.g., by 
deformation) because of the high energy barrier at the maximum of $qF(\rho )
/V.$ Hence, the dislocation density, as a function of temperature, is
\beq
\rho (T)=\left[ 
\begin{array}{cc}
0, & T<T_m, \\
\rho (T_m), & T=T_m.
\end{array}
\right.
\eeq
In fact, it can be shown that (30) is the only physical solution of (12) 
written as a ``gap'' equation: 
$$e^{-\mu /k_BT}b^2\rho (T)=A(q,z')\sum _{n=4}^\infty \frac{(z')^n}{n^q}\left( 
\frac{4b^2\rho (T)}{\alpha ^2}\right) ^{\kappa Gb^3n/8\pi k_BT}\!\!\!.$$ 
%

\begin{center}
\vspace{1.5cm} 
\epsfig{file=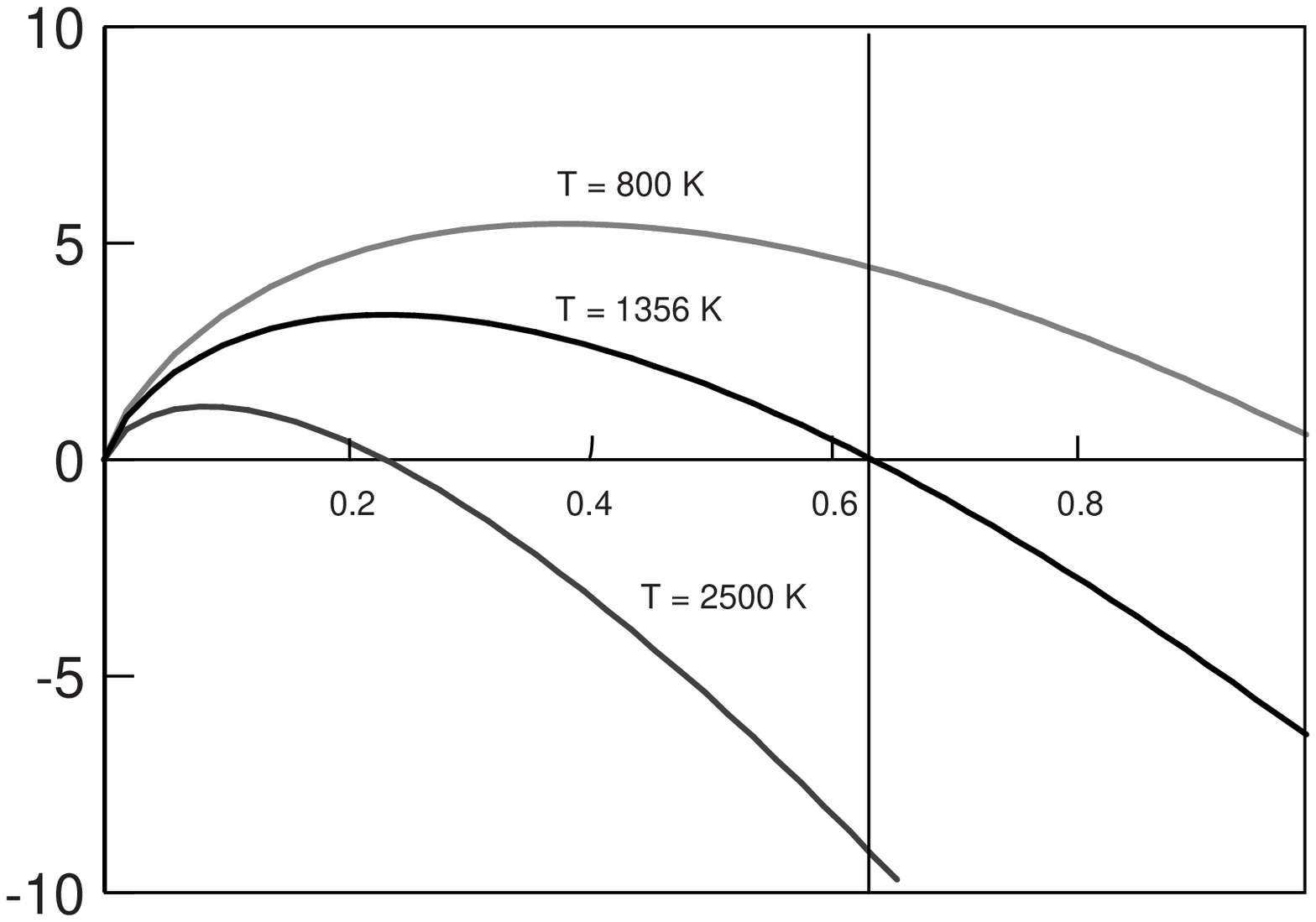,width=15cm,angle=0}
\end{center}
\vspace*{-3.7cm}
Fig.\ 2. $qF(\rho )/V$ for Cu at three different temperatures, in units 
of meV $\stackrel{\circ }{{\rm A}}$$^{-3}.$ The vertical line denotes 
the critical dislocation density value of $0.64\;\!b^{-2}.$
 \\

\section{Latent heat of fusion}

For the ensemble of strings on a lattice considered in Section 2, 
the internal energy and pressure are
\beq 
U=-\left( \frac{\partial \ln Z}{\partial \beta }\right) _{e^{\mu /k_BT},V}
\!\!\!=-\frac{\bar{N}}{Z_1}\;\!\frac{\partial Z_1}{\partial \beta }=
e^{\mu /k_BT}\;\!\frac{V\sigma }{b^2}\;\!\xi (T), 
\eeq
\beq
P=k_BT\left( \frac{\partial \ln Z}{\partial V}\right) _{T,\mu }=k_BT\frac{
\bar{N}}{Z_1}\;\!\frac{\partial Z_1}{\partial V}=e^{\mu /k_BT}\;\!\frac{
k_BT}{b^3}\;\!\bar{\xi }(T). 
\eeq
Hence, the enthalpy is 
\beq
H=U+PV=\frac{V}{b^3}\;\!e^{\mu /k_BT}\left[ \sigma b\;\!\xi (T)+k_BT\;\bar{
\xi }(T)\right] .
\eeq
The latent heat of fusion is the enthalpy difference:
\beq
L_m\equiv H(T_m)-H(0).
\eeq
In our case, $H(0)=0,$ which follows directly from (30)-(33) and Eq.\ (18). 
Using $\sigma _{{\rm eff}}(T_m)=0$ and the melting condition 
$k_BT_m=\sigma b/\ln (z-1),$ we obtain 
\beq
L_m=\frac{V}{b^3}\;\!e^{\mu (T_m)/k_BT_m}k_BT_m\left[ \ln (z-1)\xi (T_m)+
\bar{\xi }(T_m)\right] .
\eeq
To obtain the latent heat per mole, the quantity tabulated in the literature, 
one has to multiply the expression (35) by the ratio of the number of atoms 
per mole, $N_A,$ to the total number of atoms in the volume $V,$ which is 
equal to $V/v_{WS}.$ Replacing $N_Ak_B$ by the gas constant $R,$ and using 
$b^3=\lambda v_{WS},$ we obtain
\beq
L_m=\frac{e^{\mu (T_m)/k_BT_m}}{\lambda }\;\!\xi (T_m)R\;\!T_m\ln (z-1)
\left[ 1+\frac{1}{\ln (z-1)}\;\!\frac{\bar{\xi }(T_m)}{\xi (T_m)}\right] .
\eeq

To estimate the ratio $\bar{\xi }(T_m)/\xi (T_m),$ we replace 
the sums in Eqs.\ (12),(13) by the corresponding integrals:
\beq
\frac{\bar{\xi }(T_m)}{\xi (T_m)}=\frac{\int _{L/b=4}^\infty d(L/b)\;\!
(L/b)^{-q-1}}{\int _{L/b=4}^\infty d(L/b)\;\!(L/b)^{-q}}=\frac{q-1}{4q}.
\eeq
With $3/2\leq q\leq 7/4,$ as discussed in Section 2, 
$0.083\leq (q-1)/4q\leq 0.107,$ i.e., 
\beq
\frac{\bar{\xi }(T_m)}{\xi (T_m)}=0.095\pm 0.012.
\eeq
Therefore, the contribution of the second bracketed term on the right-hand 
side of Eq.\ (36) (corresponding to the work contribution to enthalpy) is 
$0.04-0.06$ $(6\leq z\leq 12).$ We expect, therefore, that neglecting the 
second bracketed term on the right-hand side of Eq.\ (36) will introduce an 
error not larger than $\sim 6$\%. Hence, with accuracy of $\sim 94$\% 
we have the following formula for the latent heats of the elements:
\beq
L_m=\frac{1}{\lambda }\;\!b^2\rho (T_m)R\;\!T_m\ln (z-1), 
\eeq
where we have replaced $e^{\mu (T_m)/k_BT_m}\xi (T_m)$ by $b^2\rho (T_m),$ in 
view of (18). The proportionality of latent heat of fusion to the critical 
concentration of defects (multiplied by the core energy) has been noted 
previously by Cotterill \cite{Cotterill}.

In Fig.\ 3 we plot the values of $b^2\rho (T_m)$ extracted from the 
experimental data on latent heats for 75 elements. For this analysis, the 
values of both $T_m$ and $L_m$ are mostly taken from \cite{Gschn}. For Be, Hf,
Sc, Sr, Y, the lanthanides Dy, Ce, Er, Gd, Ho, La, Nd, Sm, Tb, Yb, and the 
actinides Am, Cm, Th, we disregard their high-$T$ bcc phases which exist 
only in the very vicinity of melting. (The intermediate hcp$\rightarrow $fcc 
phase transition for Yb, dhcp$\rightarrow $ fcc for Am, Ce and La, and 
fcc$\rightarrow $hcp for Sr, as well as hcp$\rightarrow $fcc for Co, do not 
change coordination number.) The crystal structure chosen for the evaluation 
of Ca, Co, Mn, N, Np, O, Sm, Ti, Tl, U and Zr corresponds to the phase from 
which melting occurs. The data on both $T_m$ and $L_m$ for H, N, O, Pa and Rn 
are taken from \cite{Emsley}. The data on both $T_m$ and $L_m$ for Am and Cm, 
and on $L_m$ for Ar, Kr, Ne and Xe are taken from \cite{Tonkov}. The data on 
$L_m$ for the lanthanides are taken from \cite{Samsonov}. The following values 
of $\lambda $ are used: 1 for sc, 1.3 for bcc, 1.41 for fcc, 1.24 for Zn, 1.42 
for Mg, and 1.33 for all other elements.

\begin{center}
\vspace{2cm} 
\epsfig{file=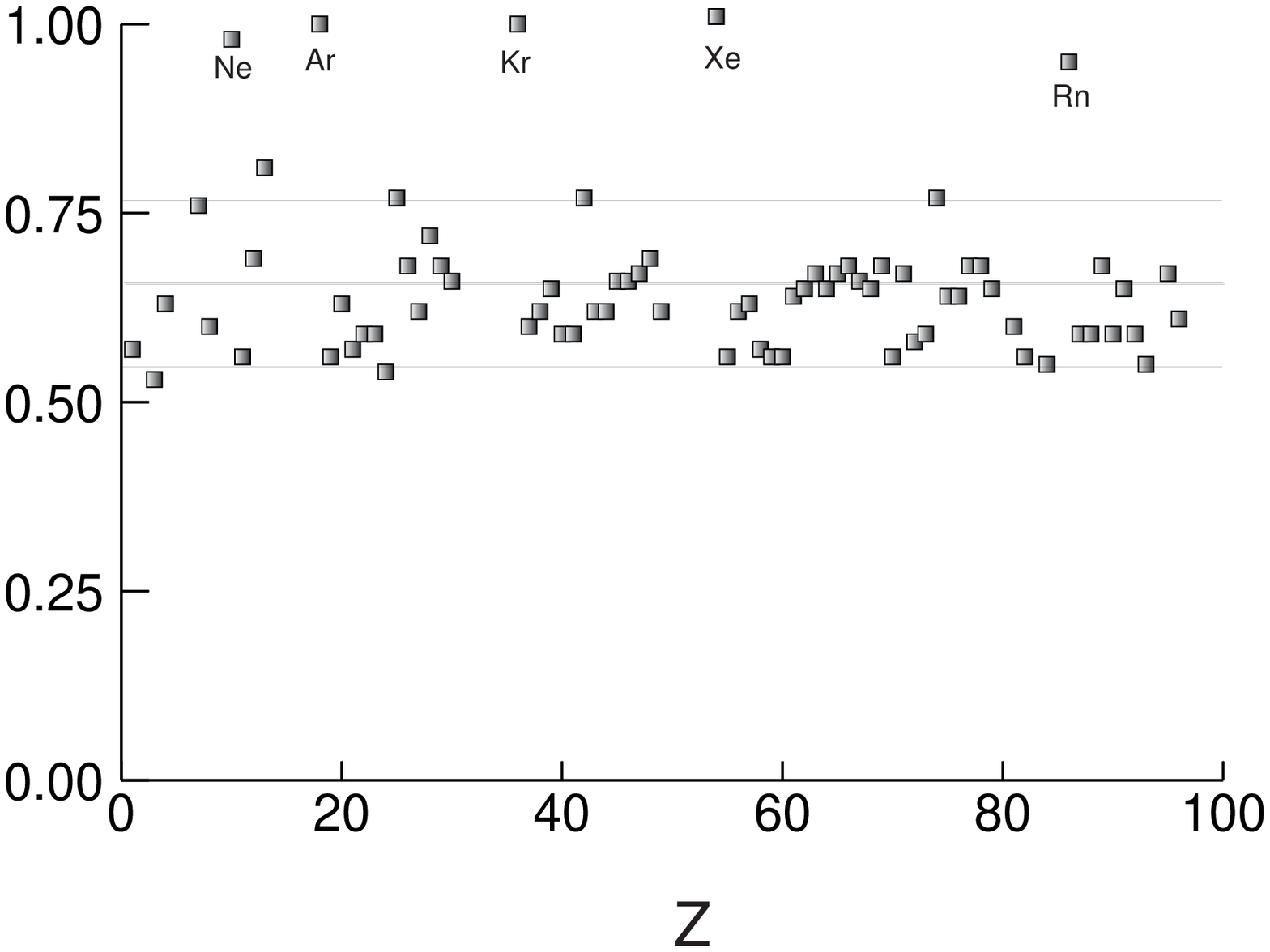,width=15cm,angle=0}
\end{center}
Fig.\ 3. Critical dislocation density as extracted from the 
experimental data on latent heat of fusion for 75 elements.
 \\

For all these elements we find
\beq
\rho (T_m)=(0.66\pm 0.11)\;\!b^{-2}, 
\eeq
where the error is the root-mean-square deviation. This value is in good 
agreement with that obtained from the melting temperatures alone, Eq.\ (23). 

Note that the possible inaccuracy in the value of $\lambda $ for the hcp, 
dhcp and bct elements used in this analysis, on the order of $\sim 7$\%, may 
slightly increase uncertainty in the value of $\rho (T_m)$ in Eqs.\ (23) and 
(40). 

We do not have a reasonable explanation for the anomalously high values of 
$\rho (T_m)$ for the noble gases. If the noble gases 
are excluded from the analysis, then $\rho (T_m)$ turns out to be 
$(0.63\pm 0.06)\;\!b^{-2}$ for the remaining 70 elements. Note also that 
for deuterium (D), which is not included in Fig.\ 3, with the data on 
$T_m$ and $L_m$ from \cite{Tonkov}, we obtain $b^2\rho (T_m)=0.70.$

The uncertainty-weighted average of the values of $\rho (T_m)$ given in 
Eqs.\ (23) and (40) is
\beq
\rho (T_m)=(0.64\pm 0.14)\;\!b^{-2}, 
\eeq
which we take as our result for the critical dislocation density at melt. 

\section{Volume change at melt} 

As an independent consistency check on the relations (23) and (40), we 
determine the critical dislocation density using the formula \cite{PP,Nin}
\beq
\varepsilon \equiv \frac{\triangle V}{V}=\frac{\lambda }{2\pi }\;\!\frac{G}{B}
\;\!\left( \gamma _G-\frac{1}{3}\right) b^2\rho (T_m),
\eeq
where $\triangle V$ is the difference between the liquid and solid specific 
volumes at melt, $G$ and $B$ are the shear and bulk moduli, respectively, and 
$\gamma _G$ is the Gr\"{u}neisen constant. Here, $\varepsilon $ is identified 
with the dilation of the lattice as the reaction of the crystal to the sudden 
proliferation of dislocations.
In Table 1 we show the values of $b^2\rho (T_m)$ calculated for 32 elements 
for which we could find zero-pressure data on $\gamma _G$ and $\varepsilon .$ 
The experimental values of $\varepsilon $ are mostly taken from ref.\ 
\cite{Tonkov}, and those of $G,$ $B$ and $\gamma _G$ from \cite{GS}. For 
Ar, Kr, Ne and Xe, the values of $G$ and $B$ are taken from \cite{KL}, 
and those of $\gamma _G$ from \cite{Vor}. 
%

\begin{center}
{\footnotesize
\begin{tabular}{|c||c|c|c|c||c||c|}
\hline 
element & $B,$ GPa & $G,$ GPa & $\gamma _G$ & $\varepsilon $ & $b^2\rho (T_m)$ 
from Eq. (42) & $b^2\rho (T_m)$ from Eq. (39)    \\
\hline 
 Ag & 103  & 29.8 & 2.40 & 0.052 & 0.39 & 0.67       \\ 
 Al & 76.0 & 26.1 & 2.19 & 0.064 & 0.44 & 0.81       \\ 
 Ar & 1.83 & 0.75 & 2.59 & 0.144 & 0.69 & 1.00       \\ 
 Au & 173  & 28.0 & 2.99 & 0.055 & 0.57 & 0.65       \\ 
 Be & 111  &  151 & 1.11 & 0.115 & 0.51 & 0.63       \\ 
 Ca & 16.7 & 7.4  & 1.15 & 0.048 & 0.64 & 0.63       \\ 
 Cs & 2.01 & 0.65 & 1.41 & 0.026 & 0.36 & 0.56       \\ 
 Cu & 137  & 47.7 & 2.02 & 0.046 & 0.35 & 0.68       \\ 
 Eu & 17.0 & 7.53 & 1.39 & 0.048 & 0.50 & 0.67       \\ 
 Gd & 37.8 & 21.6 & 0.63 & 0.021 & 0.59 & 0.65       \\ 
 Ho & 40.8 & 26.3 & 1.18 & 0.075 & 0.65 & 0.66       \\ 
 In & 42.0 & 4.78 & 2.43 & 0.025 & 0.49 & 0.62       \\ 
 K  &  3.3 & 0.9  & 1.29 & 0.025 & 0.46 & 0.56       \\ 
 Kr & 2.04 & 0.85 & 2.64 & 0.151 & 0.70 & 1.00       \\ 
 Li & 12.1 & 3.85 & 0.92 & 0.016 & 0.41 & 0.53       \\ 
 Lu & 47.6 & 27.2 & 1.06 & 0.036 & 0.41 & 0.67       \\ 
 Na & 6.74 & 1.98 & 1.19 & 0.027 & 0.52 & 0.56       \\ 
 Nb & 171  & 37.6 & 1.77 & 0.029 & 0.44 & 0.59       \\ 
 Nd & 32.9 & 17.4 & 0.57 & 0.009 & 0.34 & 0.56       \\ 
 Ne & 0.88 & 0.40 & 2.79 & 0.156 & 0.62 & 0.98       \\ 
 Ni & 183  & 85.8 & 1.93 & 0.063 & 0.37 & 0.72       \\ 
 Pb & 44.7 & 8.6  & 2.74 & 0.037 & 0.36 & 0.56       \\ 
 Pd & 193  & 48.0 & 2.56 & 0.059 & 0.47 & 0.66       \\ 
 Pt & 283  & 63.7 & 2.87 & 0.066 & 0.51 & 0.68       \\ 
 Rb & 2.3  & 0.63 & 0.99 & 0.026 & 0.70 & 0.60       \\ 
 Ta & 193  & 69.0 & 1.74 & 0.052 & 0.50 & 0.59       \\ 
 Tb & 38.7 & 22.1 & 0.74 & 0.032 & 0.65 & 0.67       \\ 
 Tl & 35.7 & 5.4  & 2.10 & 0.033 & 0.60 & 0.60       \\ 
 Tm & 46.2 & 29.1 & 1.43 & 0.069 & 0.47 & 0.68       \\ 
  W & 310  & 160  & 1.67 & 0.090 & 0.63 & 0.77       \\ 
 Xe & 2.1  & 1.0  & 2.56 & 0.130 & 0.54 & 1.01       \\ 
 Yb & 14.9 & 8.06 & 1.04 & 0.036 & 0.44 & 0.56       \\ 
\hline 
\end{tabular}
}
\end{center}
\vspace*{0.1cm}
Table 1. Values of $b^2\rho (T_m)$ from experimental data on volume change at 
melt for 32 elements. For comparison, we also show values of $b^2\rho (T_m)$ 
extracted for the same elements from the data on latent heats.
 \\
 
For all 32 elements in Table 1 we find 
\beq
\rho (T_m)=(0.51\pm 0.11)\;\!b^{-2},
\eeq
where the error is the root-mean-square deviation. This is somewhat lower 
than but still in agreement with both Eqs.\ (23) and (40) taking into account 
uncertainties associated with the three values. 

For comparison, we show in the last column of Table 1 the values of $b^2\rho (
T_m)$ extracted for the same elements from the data on $L_m.$ It is seen that 
the agreement between two sets of the values of $b^2\rho (T_m)$ is reasonably 
good, except for Ag, Al, Cs, Cu, Ni, Pb and Xe, for which the difference in 
both values of $b^2\rho (T_m)$ is on the order of $\sim 50-60$\%, Lu, Ne and 
Nd for which the difference is $\sim 45$\%, and Ar, Kr, Pd and Tm, for which 
it is $\sim 35$\%. For all other elements, the difference does not exceed 
$\sim 30$\%. 

Note that the contribution of the volume change at melt, $\varepsilon ,$ 
to the latent heat of fusion is proportional to $\varepsilon ^2\ll 1$ 
\cite{PP,Nin}, and is therefore negligibly small compared to the right-hand 
side of Eq.\ (39). 

\section{Concluding remarks}

Our theory of dislocation-mediated melting was developed in the approximation 
that dislocations are non-interacting. This approximation is good only in the 
vicinity of melt where the dislocation density is very high and the otherwise 
long-range interactions are sufficiently screened \cite{Sar}. The statistical 
mechanics of non-interacting dislocations on a lattice yields simple, accurate 
relations between the dislocation density at melt and both the melting 
temperature and latent heat of fusion, despite the indeterminacy of the 
parameter $q$ that takes into account the possible non-Brownian nature of 
the dislocation network. The values of $\rho (T_m),$ as determined from an 
extensive analysis of $T_m$ and $L_m$ data, are remarkably consistent: 
$(0.61\pm 0.20)\;\!b^{-2}$ and $(0.66\pm 0.11)\;\!b^{-2},$ respectively. The 
uncertainty-weighted average of these values is $\rho (T_m)=(0.64\pm 0.14)
\;\!b^{-2},$ which we take as our result for the dislocation density at melt. 
Poirier and Price \cite{PP} analyzed 14 elements and found $\rho (T_m)v_{
WS}/b=0.48\pm 0.12.$ Using $v_{WS}=b^3/\lambda $ with $\lambda \approx 4/3,$ 
their result corresponds to $\rho (T_m)=(0.64\pm 0.16)\;\!b^{-2},$ which is 
in excellent agreement with ours. Kierfeld and Vinokur \cite{KV} modelled  
dislocation-mediated phase transitions of a vortex-line lattice and found 
$\rho (T_m)\approx 0.6\;\!b^{-2}.$ Vachaspati's \cite{Vach} study of 
topological defect formation gave $a^2\rho (T_m) \approx 0.88$ for a simple 
cubic lattice. This translates into $\rho (T_m)\approx 0.66\;\!b^{-2}$ for 
bcc lattices $(a=2/\sqrt{3}\;b)$ and $\rho (T_m)\approx 0.44\;\!b^{-2}$ for 
fcc lattices $(a=\sqrt{2}\;\!b),$ which are consistent with our result. In 
agreement with Vachaspati, Kibble \cite{Kibble} found $a^2\rho (T_m)\approx 
0.89$ for a simple cubic lattice. 

Although our main results do not depend on the precise value of $q,$ there is 
a particular value of $q$ at which the relations (29) and (39) become exact: 
$q=1.$ In this limit, as seen in (12), $\xi (T_m)\rightarrow \infty ,$ so that 
Eq.\ (39) becomes exact in view of (36). Requiring finite internal energy in 
this limit leads, via (31), to $\exp\{\mu (T_m)/k_BT_m\}\rightarrow 0$ 
$(\mu (T_m)\rightarrow -\infty ),$ and therefore, Eq.\ (29) becomes exact, 
since the second term on the right-hand side of (28) disappears. In fact, 
the study of cosmological networks of string loops in 3 dimensions by 
Magueijo, Sandvik and Steer \cite{MSS} results in a scale-invariant loop 
distribution of the form of Eqs.\ (9),(10) with $q+1<5/2:$ \cite{Steer} $1.9
<q+1<2.1,$ or \cite{Mag} $q+1=2.03$ (plus error bars), and so in this study 
$q\approx 1.$ Thus, it is quite possible that linear defects which correspond 
to two apparently distinct physical phenomena, namely cosmic strings and 
crystal dislocations, are of a very similar statistical-mechanical nature. 

The average total dislocation length per Wigner-Seitz cell at melt is $\rho 
(T_m)v_{WS}=b\rho (T_m)/\lambda \approx b/2,$ since $\lambda \approx 4/3.$ 
Since a Wigner-Seitz cell contains $z$ links, each of length $b/2,$ it follows 
that, on average, one of $z$ links in each Wigner-Seitz cell is covered by a 
dislocation. Since each such a link is shared between two atoms, on average, 
half of the atoms are within a dislocation core at melt. 

If we use $\rho (T_m)=0.64\;\!b^{-2},$ then to $\sim 20$\% accuracy the 
melting temperatures and latent heats are given by 
\beq
k_BT_m=\frac{Gv_{WS}}{4\pi \ln (z-1)}, 
\eeq
\beq
L_m=\frac{\ln (z-1)}{2}\;\!RT_m. 
\eeq
The accuracy of these relations depends critically on the factor of $\ln (z-1
),$ which is characteristic of a theory based on line-like degrees of freedom.

\section*{Acknowledgements}
We thank T. Goldman for valuable discussions during the preparation of this 
work. One of us (L.B.) wishes to thank J. Magueijo and D.A. Steer for very 
useful correspondence.


\bigskip
\bigskip
\begin{thebibliography}{9}
\bibitem{Shock} W. Shockley, in {\it L'Etat Solide,} Proceedings of 
Neuvienne-Consail de Physique, Brussels, 1952, Ed. R. Stoops, (Solvay Institut
de Physique, Brussels, Belgium) 
\bibitem{Bragg} W.L. Bragg, in {\it Proc. Symp. on Internal Stresses,} 
(Institute of Metals, London, 1947), p.\ 221
\bibitem{CD} R.M.J. Cotterill and M. Doyama, Phys. Rev. {\bf 145} (1966) 465 
\bibitem{Mott} C. Mott, Proc. Roy. Soc. A {\bf 215} (1952) 1
\bibitem{MD} R.M.J. Cotterill, Phys. Rev. Lett. {\bf 42} (1979) 1541 
\bibitem{MC} W. Janke, Int. J. Theor. Phys. {\bf 29} (1990) 1251 
\bibitem{Crawf} R.K. Crawford, Bul. Am. Phys. Soc. {\bf 24} (1979) 385 \\
R.M.J. Cotterill and J.K. Kristensen, Phil. Mag. {\bf 36} (1977) 453 
\bibitem{Miz} S. Mizushima, J. Phys. Soc. Japan {\bf 15} (1960) 70 
\bibitem{Ook} A. Ookawa, J. Phys. Soc. Japan {\bf 15} (1960) 2191 
\bibitem{Siol} M. Siol, Z. Phys. {\bf 164} (1961) 93 \\ 
D. Kuhlmann-Wilsdorf, Phys. Rev. {\bf 140} (1965) A1595 \\ 
S.F. Edwards and M. Warner, Phil. Mag. A {\bf 40} (1979) 257 \\ 
S.P. Obukhov, Zh. Eksp. Teor. Fiz. {\bf 83} (1982) 1978 
\bibitem{Pat} A.Z. Patashinskii and B.I. Shumilo, Zh. Eksp. Teor. Fiz. 
{\bf 89} (1985) 315 \\ 
A.Z. Patashinskii and L.D. Son, Zh. Eksp. Teor. Fiz. {\bf 103} (1993) 1087 
\bibitem{Copeland2} E. Copeland, D. Haws, S. Holbraad and R. Rivers, 
Physica A {\bf 179} (1991) 507  
\bibitem{PP} J.P. Poirier and G.D. Price, Phys. Earth Planet. Inter. {\bf 69}
(1992) 153 
\bibitem{KleinII} H. Kleinert, {\it Gauge Fields in Condensed Matter,} (World
Scientific, Singapore, 1989), Vol.\ II, and references therein 
\bibitem{Cotterill} R.M.J. Cotterill, J. Cryst. Growth {\bf 48} (1980) 582 
\bibitem{reviews} B.I. Halperin, Statistical mechanics of topological defects, 
in {\it Physics of Defects,} Eds. R. Balian, M. Kleman and J.P. Poirier, 
(North Holland, Amsterdam, 1981) \\ 
J.P. Poirier, Geophys. J. Roy. Astr. Soc. {\bf 85} (1986) 315 
\bibitem{YI} T. Yamamoto and T. Izuyama, J. Phys. Soc. Japan {\bf 57} (1988) 
3742 
\bibitem{KV} J. Kierfeld and V. Vinokur, Dislocations and the critical 
endpoint of the melting line of vortex line lattices, cond-mat/9909190 
\bibitem{previous} L. Burakovsky and D.L. Preston, Los Alamos preprint 
LA-UR-99-4171 [cond-mat/0003494], to appear in Solid State Comm. 
\bibitem{Wie} F.W. Wiegel, {\it Introduction to Path-Integral Methods in 
Physics and Polymer Science,} (World Scientific, Singapore, 1986)
\bibitem{GSW} M.B. Green, J.H. Schwarz and E. Witten, {\it Superstring 
Theory,} (Cambridge University Press, Cambridge, 1987), Vol.\ 1 
\bibitem{Copeland} E. Copeland, D. Haws, S. Holbraad and R. Rivers, 
Physica A {\bf 158} (1989) 460, Nucl. Phys. B {\bf 319} (1989) 687  
\bibitem{BV} R. Brandenberger and C. Vafa, Nucl. Phys. B {\bf 316} (1989) 391 
\bibitem{SS} P. Salomonson and B.S. Skagerstam, Nucl. Phys. B {\bf 268} (1986) 
349, Physica A {\bf 158} (1989) 499  
\bibitem{LT} D.A. Lowe and L. Thorlacius, Phys. Rev. D {\bf 51} (1995) 665 
\bibitem{HL} J.P. Hirth and J. Lothe, {\it Theory of Dislocations,} 2nd ed.,
(Krieger Publishing, Malabar, FL, 1992)
\bibitem{Sar} G.F. Sarafanov, Phys. Solid State {\bf 39} (1997) 1403 
\bibitem{GS} M.W. Guinan and D.J. Steinberg, J. Phys. Chem. Solids {\bf 35}
(1974) 1501
\bibitem{dipole} N.D. Mermin, Rev. Mod. Phys. {\bf 51} (1979) 591 \\ 
N.K. Gilra, Cryst. Lattice Defects {\bf 8} (1979) 59 \\
K.J. Strandburg, Rev. Mod. Phys. {\bf 60} (1988) 161
\bibitem{KT} J.M. Kosterlitz and D.J. Thouless, J. Phys. C {\bf 6} (1973) 1181,
in {\it Progress in Low Temperature Physics,} Vol. VII-B, Ed. D.F. Brewer, 
(North-Holland, Amsterdam, 1978), p. 373. The $3D$ Kosterlitz-Thouless-like 
transition is discussed in C. Giannessi, J. Phys. Cond. Mat. {\bf 3} (1991) 
1649, and N.K. Kultanov and Yu.E.~Lozovik, Solid State Comm. {\bf 88} (1993) 
645, Physica Scripta {\bf 56} (1997) 129  
\bibitem{HN} D.R. Nelson and B.I. Halperin, Phys. Rev. Lett. {\bf 41} (1978) 
121, Phys. Rev. B {\bf 19} (1979) 2456 
\bibitem{NT} D.R. Nelson and J. Toner, Phys. Rev. B {\bf 24} (1981) 363 
\bibitem{Gschn} K.A. Gschneidner, Jr., in {\it Solid State Physics, Advances 
in Research and Applications,} Eds. F. Seitz and D. Turnbull, (Academic Press,
New York, 1965), Vol. 16, p.~275
\bibitem{Emsley} J. Emsley, {\it The Elements,} (Clarendon Press, Oxford, 1989)
\bibitem{Tonkov} E. Yu. Tonkov, {\it High Pressure Phase Transformations,} 
(Gordon and Breach, Philadelphia, 1992)
\bibitem{Samsonov} {\it Handbook of the Physicochemical Properties of the 
Elements,} Ed. G.V. Samsonov, (IFI/Plenum, New York, 1968)
\bibitem{Nin} T. Ninomiya, J. Phys. Soc. Japan {\bf 44} (1978) 263
\bibitem{KL} P. Korpiun and E. L\"{u}scher, in {\it Rare Gas Solids,} Eds. 
M.L. Klein and J.A. Venables, (Academic Press, London, 1977), Vol. II, p. 741
\bibitem{Vor} V.S. Vorob'ev, High Temp. {\bf 34} (1996) 197
\bibitem{Vach} T. Vachaspati, Phys. Rev. D {\bf 44} (1991) 3723 
\bibitem{Kibble} T.W.B. Kibble, Phys. Lett. B {\bf 166} (1986) 311 
\bibitem{MSS} J. Magueijo, H. Sandvik and D.A. Steer, Phys. Rev. D {\bf 60} 
(1999) 103514 
\bibitem{Steer} D.A. Steer, private communication 
\bibitem{Mag} J. Magueijo, private communication  
\end{thebibliography}
\end{document}